\documentclass[pra,aps,amssymb,amsmath,amsmath,showpacs,reprint,twocolumn]{revtex4-1}
\usepackage{color}
\usepackage{graphicx}
\usepackage{epstopdf}
\usepackage{natbib}
\usepackage{hyperref}
\usepackage{dcolumn}
\usepackage{bm}
\usepackage{dcolumn}
\newcolumntype{d}[1]{D{.}{.}{#1}}

\newcommand{\Fkt}[1]{\,\mathsf {#1}}
\def\openone{\leavevmode\hbox{\small1\kern-3.3pt\normalsize1}}

\ifx\Tr\renewcommand{\Tr}{\Fkt{Tr}} 
\else\newcommand{\Tr}{\Fkt{Tr}}
\fi

\usepackage{booktabs}
\usepackage{graphicx}
\usepackage{amsmath}
\usepackage{indentfirst}
\begin{document}
\title{Theoretical determination of polarizability and magnetic susceptibility of neon}

\author{\sc Micha\l\ Lesiuk}
\email{e-mail: lesiuk@tiger.chem.uw.edu.pl}
\author{\sc Micha\l\ Przybytek}
\author{\sc Bogumi\l\ Jeziorski}
\affiliation{\sl Faculty of Chemistry, University of Warsaw\\
Pasteura 1, 02-093 Warsaw, Poland}
\date{\today}

\begin{abstract}
We report theoretical determination of the dipole polarizability of the neon atom, including its frequency dependence. 
Corrections for the relativistic, quantum electrodynamics, finite nuclear mass, and finite nuclear size effects are 
taken into account. We obtain the value 
$\alpha_0=2.66080(36)$ for the static polarizability, and $\alpha_2=2.850(7)$ and 
$\alpha_4=4.932(14)$ for the first two polarizability dispersion coefficients (Cauchy moments); all values are in atomic 
units (a.u.).  In 
the case of static polarizability, our result agrees with the best experimental determination  [C. Gaiser and B. 
Fellmuth, Phys. Rev.
Lett. \textbf{120}, 123203 (2018)],  but our  estimated uncertainty is  significantly  larger. 
For the dispersion coefficients, the results obtained in 
this work appear to be the most accurate to date overall compared to published theoretical 
and experimental data. We also calculated the static   magnetic susceptibility  of the neon atom, 
needed  to obtain the refractive index of gaseous neon. Our result, $\chi_0 = -8.484(19) \cdot 
10^{-5}$ a.u., is about  9\% larger in absolute value  than  the  recommended  experimental value  
[CRC Handbook of Chemistry and Physics,  CRC Press, 2019, p. 4-145].   
\end{abstract}

\maketitle

\section{Introduction}
\label{sec:intro}

Electric dipole polarizability, $\alpha$, is one of the fundamental properties of atomic and molecular systems as it  
determines response to a perturbation by an external electric field. The importance of polarizability manifests itself 
also through the Clausius-Mossotti equation 
\begin{align}
\label{cmr}
 \frac{\epsilon_r-1}{\epsilon_r+2} = \frac{4\pi}{3}\, \alpha\rho,
\end{align}
which connects the relative electric permittivity  $\epsilon_r$  of an atomic gas at low densities  
$\rho$ with an atomic property. 
A formula analogous to Eq.~(\ref{cmr}) holds also for the second important quantity -- relative magnetic 
permeability, $\mu_r$. In this case the polarizability in Eq.~(\ref{cmr}) is replaced by the magnetic 
susceptibility of the gas, $\chi$. Together, the values of $\epsilon_r$ and $\mu_r$ determine the 
refractive index, $n=\sqrt{\epsilon_r\,\mu_r}$, which is fundamental in determination of optical 
properties of materials. The relation~(\ref{cmr}) is valid only in the low-density limit, 
but corrections to this equation proportional to higher powers of $\rho$ (expressed through the so-called virial 
coefficients) are relatively small for noble gases \cite{song20}.

The connection between the microscopic quantities, $\alpha$ and $\chi$, and the macroscopic ones, 
$\epsilon_r$ and $\mu_r$, becomes particularly important if one notices that the latter two are directly accessible 
by experimental techniques.
The relative electric permittivity can be determined from capacitance measurements 
\cite{fellmuth14,gaiser15,guenz17,gaiser19,gaiser20} 
with  accuracy reaching one part per million. In fact, such measurements  (dielectric-constant gas 
thermometry)
are the source  of currently most accurate experimental values of  atomic  polarizabilities  \cite{Gaiser18}.  The refractive index can be 
measured by several microwave \cite{schmidt07} and optical methods \cite{jousten17,hendricks18}. 

Development of new 
experimental techniques~\cite{fellmuth14,gaiser15,guenz17,gaiser19,gaiser20,Gaiser18,schmidt07,
jousten17, hendricks18} has been critical for the progress in modern thermal metrology.  For 
example, it has been 
proposed to establish a new pressure standard based on  the  measurements of $\epsilon_r$,  or  on laser refractometry 
measurements of the 
refractive index \cite{hendricks18}. The 
link between pressure $p$ of the gas and the refractive index $n$ is provided by the following equation
\begin{align}
 p=\frac{kT}{2\pi(\alpha+\chi)}(n-1),
\end{align}
valid in the low-pressure limit. Since the value of the Boltzmann constant $k$ is now fixed according to the new SI 
definition of the unit of temperature $T$  \cite{fischer19,machin19}, the 
knowledge of $\alpha$ and $\chi$ is required to obtain $p$ directly from $n$, via Eq. 
(2).

A significant portion of the experimental effort in this field targets noble gases due to numerous favorable 
properties such as being chemically inert and stable, etc. The simplest member of this family, i.e., helium, has been 
intensively studied in recent years also by theoretical methods 
\cite{johnson96,bhatia98,pachucki00,cencek01,lach04,puchalski11,piszczu15,puchalski16}. The static 
dipole polarizability of the helium atom is 
now known from theory with relative accuracy of  $10^{-7}$ \cite{puchalski20}. This is more 
than sufficient for the purposes of 
metrology. However, the same cannot be said about the heavier noble gas atoms. For example, most systematic theoretical 
studies of the polarizability of the neon atom were undertaken more than a decade 
ago~\cite{klopper04,larsen99,franke01,hald03,pawlowski05,chernov05,hammond08,kumar10}. Additionally, if one aims at the 
accuracy better than one part per thousand it is necessary to include various corrections accounting 
for effects 
beyond the clamped-nuclei nonrelativistic Schr\"{o}dinger equation which has not been systematically done thus far. 

The main purpose of the present work is state-of-the-art theoretical determination of the static 
and dynamic polarizability of the neon atom at experimentally relevant frequencies. To this end, we 
employ high-level \emph{ab initio} electronic structure methods and basis sets designed 
specifically for the task. Our study is not limited to the nonrelativistic picture and we consider corrections that 
account for other relevant physical effects. This includes relativistic, quantum electrodynamics (QED), finite nuclear 
mass, and finite nuclear size effects.

We also compute the  magnetic susceptibility  $\chi$ relevant for the measurements of the refractive index.  Since 
$\chi$  is about five orders of magnitude smaller than $\alpha$, it does not have to be known with very high accuracy 
when used in Eq. (2).  Therefore, we determine   $\chi$  
using the nonrelativistic theory. An estimate of the neglected relativistic and QED effects is
included in the final uncertainty estimate. 
 
Atomic units  (a.u.) are used throughout the present work unless explicitly stated otherwise. The following values of physical 
constants are assumed: speed of light in vacuum, $c=137.035\,999\,084 $, proton-to-electron mass ratio, 
$ m_{\rm p}/m_{\rm e} = 1836.152\,673\,43$, Bohr radius, $ a_0 = 0.529\,177\,210\,903\,$\AA{}, Avogadro number, 
$N_A=6.022\,140\,76 \cdot10^{23}$, according to the most recent CODATA database~\cite{codata18}. All results given in 
this paper refer to the most abundant $^{20}$Ne isotope of neon with the nuclear mass equal 
to~$36434.0\, m_{\rm e}$ \cite{wang17}.

\section{Theoretical calculations}

\subsection{Clamped-nucleus nonrelativistic polarizability}

Polarizability is a quantity that depends on the frequency of the external electric field ~$\omega$  that perturbs 
the system. This frequency dependence is non-negligible in accurate treatments and must be taken into account in 
calculation of quantities appearing in Eq.~(\ref{cmr}). Assuming that the necessary range of frequencies corresponds to 
energies well below the first atomic resonant frequency~\cite{saloman04} 
($\omega\approx0.6107\,$a.u), we can use the following expansion
\begin{align}
\label{poldisp}
 \alpha(\omega) = \alpha_0 + \alpha_2\,\omega^2 + \alpha_4\,\omega^4 + \ldots
\end{align}
where $\alpha_2$, $\alpha_4$, $\ldots$ are the so-called dispersion coefficients (Cauchy coefficients) and 
$\alpha_0:=\alpha(0)$ is a shorthand notation for the static polarizability. 

For frequencies that are of particular experimental interest, e.g., helium-neon laser operating at wavelengths near 
$633\,$nm (which corresponds to~$\omega\approx0.072\,$ a.u.), the expansion~(\ref{poldisp}) is rapidly 
convergent. The inclusion of terms only up to $\omega^4$ is sufficient to reach relative accuracy in~$\alpha(\omega)$ 
better than~$10^{-6}$ for the neon atom at this frequency. By retaining only the quadratic term in Eq.~(\ref{poldisp}), 
the~$10^{-4}$ accuracy level is attainable. Moreover, assuming that $\alpha_0$ was calculated with relative accuracy of 
about~$10^{-5}$ and the same accuracy level is to be retained in~$\alpha(\omega)$, the coefficient~$\alpha_2$ must be 
accurate to a few parts per thousand, and the coefficient~$\alpha_4$ to only about~10\%. This greatly simplifies the 
incorporation of frequency dependence into the theoretical results.

The first step in the calculations is an accurate determination of the nonrelativistic polarizability of the neon atom 
with inclusion of the frequency-dependence, i.e. $\alpha_0$, $\alpha_2$, and $\alpha_4$. These calculations were 
performed using the basis sets of Slater-type orbitals  (STO) \cite{slater30,slater32} optimized 
specifically for the purposes of the present work. 
The basis sets are designated wtcc$X$ (well-tempered correlation-consistent) where~$X=2,\ldots,7$, referred  usually to as the cardinal number, stands for the 
highest angular momentum $l$ included in the basis. 
They are  designed according to the  correlation-consistency principle~\cite{dunning89}
 which allows for a reliable extrapolation towards the complete basis set limit (the basis wtcc$2$ 
has the composition $6s4p1d$ and each consecutive  basis has one more function for each 
angular momentum $l\le X$).  To properly allow for the polarization by the electric field, these 
bases are augmented  by adding  diffuse functions  (with small exponents).  
The  notation ``no-aug'' means 
no augmentation,  while  ``s-aug'',  ``d-aug'', ``t-aug'',  and``q-aug''  stand for the inclusion of one,  two, three, and four diffuse functions  for each angular momentum. 
These  bases, optimized using the methodology established in Refs.~\cite{lesiuk15,lesiuk17}, are  
available from authors 
upon request.  Computer codes for the STO integrals, described in Refs.~\cite{lesiuk14a,lesiuk14b,lesiuk16}, were used 
in all relevant calculations. Most calculations reported in this paper were accomplished using variants of the coupled 
cluster (CC) theory; for a detailed review of this family of methods see Ref.~\cite{bartlett07}.

To test the quality of the basis sets developed in this work, the static polarizability of the helium atom was 
calculated. The optimization and composition of the helium basis sets is fully analogous to the 
ones described above - the only difference is that the number of the $s$ functions is smaller by 
one at each 
augmentation and polarization level. The helium atom constitutes an excellent benchmark for the purposes of this work 
since a practically exact value of the non-relativistic clamped-nuclei static polarizability is 
known,~$\alpha_0(\mbox{He})=1.383\,192\,174\,455(1)$, as reported by Pachucki and Sapirstein 
\cite{pachucki00}. Various corrections beyond this level of theory were also reported in the literature~\cite{liu91,pachucki00,lach04,sahoo08,puchalski16,puchalski20}.

In Table \ref{tab:a0he}, the static polarizability of helium atom calculated using the coupled 
cluster theory with single and double substitutions~\cite{purvis82,scuseria87} (CCSD) is 
presented. 
Since for two-electron systems the CCSD method is equivalent to the full configuration interaction 
(FCI), the only source of error in 
these results (compared with the reference value cited above) is the 
incompleteness of the basis set. While the results are 
well-saturated with respect to the augmentation level, the same is not true when considering the maximum angular 
momentum $l=X$ included in the basis set. To   circumvent  this problem  and to improve the convergence with increasing  cardinal number $X$, the results were extrapolated by using the 
conventional~$X^{-3}$ formula \cite{hill85,helgaker97,halkier98}. The value extrapolated from two 
consecutive basis sets ($X$ and $X+1$) shall be denoted 
CBS($X$,$X+1$) where the abbreviation CBS stands for the complete basis set limit. For simplicity, 
extrapolation 
with respect to the augmentation level was not attempted and q-aug basis sets were used throughout. 
The results of the performed extrapolations are
\begin{align*}
 \mbox{CBS}(3,4) = 1.38337,\\
 \mbox{CBS}(4,5) = 1.38374,\\
 \mbox{CBS}(5,6) = 1.38326,\\
 \mbox{CBS}(6,7) = 1.38322.
\end{align*}
It is reasonable to assume that the error of the last result is no larger than the difference between the  
   $\mbox{CBS}(5,6)$ and $\mbox{CBS}(6,7)$ extrapolations. This gives our estimation for the non-relativistic 
clamped-nuclei static polarizability of the helium atom,~$\alpha_0(\mbox{He})=1.38322(4)$. Compared with the reference 
value given above, the true error of this result is about 2 parts per $10^5$. Moreover, 
the reference value 
lies within the error bars estimated~by~us.

\begin{table}[t]
\caption{\label{tab:a0he}
Linear-response CCSD static polarizability ~$\alpha_0$  calculated for the 
helium atom. }
\begin{ruledtabular}
\begin{tabular}{ccccccc}
       & wtcc2 & wtcc3 & wtcc4 & wtcc5 & wtcc6 & wtcc7 \\
\hline\\[-1em]
no-aug & 1.37181 & 1.37576 & 1.37886 & 1.38118 & 1.38198 & 1.38253 \\
s-aug  & 1.38402 & 1.38345 & 1.38335 & 1.38355 & 1.38344 & 1.38339 \\
d-aug  & 1.38414 & 1.38384 & 1.38354 & 1.38365 & 1.38350 & 1.38341 \\
t-aug  & 1.38401 & 1.38388 & 1.38358 & 1.38366 & 1.38350 & 1.38340 \\
q-aug  & 1.38392 & 1.38390 & 1.38359 & 1.38367 & 1.38350 & 1.38340 \\
\end{tabular}
\end{ruledtabular}
\end{table}

\begin{table}[t]
\caption{\label{tab:a0cc3}
Linear-response CC3 static polarizability ~$\alpha_0$  calculated for the neon 
atom with all electrons correlated. }
\begin{ruledtabular}
\begin{tabular}{ccccccc}
       & wtcc2 & wtcc3 & wtcc4 & wtcc5 & wtcc6 & wtcc7 \\
\hline\\[-1em]
no-aug & 0.81910 & 1.49892 & 1.60219 & 1.99140 & 2.13575 & 2.26198 \\
s-aug  & 2.43166 & 2.33238 & 2.49043 & 2.55603 & 2.58556 & 2.60764 \\
d-aug  & 2.65436 & 2.65557 & 2.67004 & 2.66564 & 2.66406 & 2.66294 \\
t-aug  & 2.66676 & 2.68160 & 2.67124 & 2.66674 & 2.66474 & 2.66362 \\
q-aug  & 2.66926 & 2.68142 & 2.67115 & 2.66678 & 2.66473 & 2.66361 \\
\end{tabular}
\end{ruledtabular}
\end{table}

Our workhorse method for determination of the non-relativistic clamped-nuclei polarizability is the orbital-unrelaxed 
linear-response CC3 method as implemented in the \textsc{Dalton} program 
package~\cite{daltonpaper,dalton18,hald03,pawlowski05}. For the purposes of 
this work, the STO integral code was interfaced with the \textsc{Dalton} program. The CC3 method is 
an advanced approximate variant of the  coupled-cluster method designed to include the major part of 
the contribution of three-electron excitations in a computationally efficient way. The 
CC3 polarizability is calculated with help of the linear response function obtained from 
time-dependent quasienergy Lagrangian. The resulting equations for the triply excited component of 
the linear response function are then truncated at the second order in the fluctuation potential 
enabling an efficient evaluation, see Ref.~\cite{hald03} for a complete presentation.

In Table~\ref{tab:a0cc3}, we present CC3 results of the static polarizability of the neon atom. The 
values converge rather quickly with the 
augmentation level, practically as fast as in the case of the helium atom. For example, in all basis sets other than the 
smallest ones, the transition from t-aug to q-aug basis sets changes the results only at the sixth significant digit. 
Therefore, the larger triply-augmented basis sets can be viewed as saturated with respect to the augmentation level. 
Unfortunately, the convergence is not as rapid with respect to the increasing angular momentum in the basis set. To  remedy 
this problem we extrapolated the results to the complete basis set limit by using the same 
strategy as for the helium atom. Extrapolations of the results from the q-aug basis sets give
\begin{align*}
 \mbox{CBS}(3,4) = 2.66365,\\
 \mbox{CBS}(4,5) = 2.66220,\\
 \mbox{CBS}(5,6) = 2.66190,\\
 \mbox{CBS}(6,7) = 2.66172.
\end{align*}
One can see that the extrapolated results are very stable and that the convergence towards the limit is markedly 
improved. While the estimation of errors shall be a subject of further discussion, observing the convergence pattern of 
the extrapolated results allows us to suggest that the value~$\alpha_0=2.66172(18)$ is a reasonable 
estimate for the basis set limit of the CC3 static polarizability of the neon atom. The estimated 
error of this result is thus about 7 parts per 10$^5$, about 3.5 times larger than in 
the case of the helium atom.

\begin{table}[b]
\caption{\label{tab:a01w}
Core contribution to the static polarizability $\alpha_0$  of the neon 
atom calculated at the CC3 level of theory.}
\begin{ruledtabular}
\begin{tabular}{ccccccc}
       & wtcc2 & wtcc3 & wtcc4 & wtcc5 & wtcc6 \\
\hline\\[-1em]
no-aug & 0.00016 & 0.00104 & 0.00165 & 0.00322 & 0.00389 \\
s-aug  & 0.00210 & 0.00325 & 0.00434 & 0.00522 & 0.00560 \\
d-aug  & 0.00295 & 0.00457 & 0.00527 & 0.00584 & 0.00607 \\
t-aug  & 0.00300 & 0.00492 & 0.00534 & 0.00587 & 0.00608 \\
q-aug  & 0.00302 & 0.00503 & 0.00537 & 0.00589 & 0.00609 \\
\end{tabular}
\end{ruledtabular}
\end{table}

Although all electrons were correlated to obtain the results presented in Table \ref{tab:a0cc3},  
it is of interest to see  what is the contribution  to $\alpha_0$ coming from the core  ($1s^2$) and 
core-valence correlation.  In Table~\ref{tab:a01w}, we show the difference between CC3 polarizabilities calculated with 
frozen 
 core and with all electrons correlated. It turns out that the effect  of the core  and core-valence  correlation  is relatively small and stabilizes 
more rapidly with the size of the basis set than the all-electron results  reported in Table~\ref{tab:a0cc3}. 
By extrapolating the contributions from Table~\ref{tab:a01w} obtained with the q-aug basis sets,  
we find 
 \begin{align*}
\mbox{CBS}(4,5) = 0.00644,\\
\mbox{CBS}(5,6) = 0.00637.
\end{align*}
so that the final estimate for the core  contribution to the static polarizability of the neon atom is~$0.00637(7)$. 
This constitutes only about~2\% of the total CC3 correlation contribution to the static polarizability. We assume it  is unlikely that this ratio increases substantially in calculation of   higher-order correlation effects and  use the  results of Table~\ref{tab:a01w}   to justify the neglect of   core contribution to  some higher-order correlation effects  discussed further in the text.

\begin{table}[t]
\caption{\label{tab:aotq}
Post-CC3 triples and quadruples corrections to the static polarizability,~$\alpha_0$, of the 
neon atom (frozen~$1s$ core orbital).}
\begin{ruledtabular}
\begin{tabular}{ccccc}
       & cc-pVDZ & cc-pVTZ & cc-pVQZ & cc-pV5Z \\
\hline\\[-1em]
& \multicolumn{4}{c}{CCSDT -- CC3 correction} \\
\hline
s-aug  & 0.00137 & 0.00212 & 0.00142 & 0.00079 \\
d-aug  & 0.00119 & 0.00119 & 0.00065 & 0.00050 \\
t-aug  & 0.00115 & 0.00109 & 0.00063 & 0.00049 \\
\hline\\[-1em]
& \multicolumn{4}{c}{full Q correction} \\
\hline
s-aug  & $-$0.00222 & $-$0.00422 & $-$0.00489 & --- \\
d-aug  & $-$0.00542 & $-$0.00597 & --- & --- \\
t-aug  & $-$0.00555 & $-$0.00597 & --- & --- \\
\end{tabular}
\end{ruledtabular}
\end{table}

The CC3 method is an approximate model which misses some of the effects of the triple excitations and neglects higher 
excitations entirely. Therefore, it is necessary to account for these effects using a higher-level 
theory. The complete post-CC3 contribution to the polarizability is split into three components
\begin{itemize}
 \item post-CC3 triples contribution, \emph{i.e.}, the difference between the 
coupled cluster method with single, double and triple excitations~\cite{noga87,scuseria88} (CCSDT) 
and the CC3 results;
 \item full quadruples contribution, \emph{i.e.}, the difference between the coupled cluster method 
with single, double, triple and quadruple excitations~\cite{kucharski92,matthews15} (CCSDTQ) and 
the CCSDT results;
 \item post-CCSDTQ contributions, \emph{i.e.}, the difference between the FCI and CCSDTQ results. 
\end{itemize}
We also explored  the 
possibility of using methods that account for the quadruples perturbatively, such as 
CCSDT[Q]~\cite{kucharski89} and CCSDT(Q)~\cite{bomble05}.
However, we found them to be very unreliable in (finite-field) calculation of properties. For this reason the use of 
perturbative quadruple models was 
abandoned. 

In  CCSDT and CCSDTQ calculations, we employed the aug-cc-pV$X$Z Gaussian basis set family,~$X=2-5$ 
\cite{dunning89,kendall92}. To saturate the results with respect to the augmentation level, we additionally 
created the doubly-augmented (d-aug-cc-pV$X$Z) and triply-augmented (t-aug-cc-pV$X$Z) variants. They were obtained by 
scaling the smallest exponent of the Gaussian function in a given shell by the ratio of the smallest two. The CCSDT and 
CCSDTQ polarizabilities were calculated analytically (without orbital relaxation) with 
help of the \textsc{MRCC} program package~\cite{kallay20}. To reduce the computational burden, we 
froze the~$1s^2$ core of the neon atom; as discussed above this approximation is accurate to 1--2\% 
and thus entirely sufficient for 
the present purposes.

\begin{table}[b]
\caption{\label{tab:a2a4cc3}
Linear-response CC3 dispersion coefficients $\alpha_2$ and $\alpha_4$ calculated for the neon 
atom with all electrons correlated.
}
\begin{ruledtabular}
\begin{tabular}{ccccccc}
       & wtcc2 & wtcc3 & wtcc4 & wtcc5 & wtcc6 & wtcc7 \\
\hline
\multicolumn{7}{c}{$\alpha_2$}\\
\hline
no-aug & 0.20491 & 0.90963 & 0.66040 & 1.52304 & 1.82238 & 1.94098 \\
s-aug  & 1.92647 & 2.00067 & 2.14567 & 2.40571 & 2.49805 & 2.57002 \\
d-aug  & 2.93317 & 2.72236 & 2.84814 & 2.84109 & 2.84239 & 2.83986 \\
t-aug  & 2.94922 & 2.90711 & 2.87749 & 2.86330 & 2.85852 & 2.85389 \\
q-aug  & 2.95149 & 2.91005 & 2.87799 & 2.86392 & 2.85861 & 2.85398 \\
\hline
\multicolumn{7}{c}{$\alpha_4$}\\
\hline
no-aug & 0.08712 & 1.07861 & 0.48510 & 2.36422 & 3.45964 & 3.52939 \\
s-aug  & 2.66424 & 3.42031 & 3.11409 & 4.02005 & 4.13486 & 4.24444 \\
d-aug  & 5.39007 & 4.57093 & 4.85203 & 4.85022 & 4.85212 & 4.85396 \\
t-aug  & 5.34708 & 5.08456 & 5.00535 & 4.96669 & 4.94596 & 4.93349 \\
q-aug  & 5.33855 & 5.11252 & 5.00519 & 4.96531 & 4.94504 & 4.93266 \\
\end{tabular}
\end{ruledtabular}
\end{table}

The full T and Q corrections to the static polarizability of the neon atom are given in Table~\ref{tab:aotq}. 
Extrapolation of the full triples correction using the results from the t-aug basis sets gives
\begin{align*}
 \mbox{CBS}(3,4) = 0.00029,\\
 \mbox{CBS}(4,5) = 0.00034,
\end{align*}
and an extrapolation with respect to the augmentation level does not seem to be necessary. This allows us to estimate 
that the full triples correction amounts to~$0.00034(5)$, where the uncertainty is the difference between the last 
two extrapolated values. Moving to the quadruples correction, the same procedure gives only one value
\begin{align*}
 \mbox{CBS}(2,3) = -0.00614,\\
\end{align*}
and our final estimation of this correction is~$-0.00614(17)$, where the error is the difference between the 
extrapolated value and the result obtained with the t-aug-cc-pVTZ basis. The results shown in 
Table~\ref{tab:aotq} 
show that 
the T and Q corrections to the static polarizability of the neon atom are of the order of a few parts per thousand, 
and thus are non-negligible from the present point of view. To 
account for 
even higher-order excitations we employ the FCI method using dedicated codes written by one of 
us~\cite{przybytek14}. This is 
very computationally expensive and 
we managed to 
calculate the FCI correction only in smaller basis sets. Since no extrapolation can be 
performed we simply add 
the value calculated with the d-aug-cc-pVDZ (the largest basis set for which the FCI result could be obtained) and 
assign 
50\% uncertainty to it, getting~$-0.00047(24)$. It is worth pointing out that even within this 
relatively small basis, the number of configurations that were included in the FCI calculations 
reached about $2\cdot10^9$.

Results of the calculations of the dispersion coefficients $\alpha_2$ and $\alpha_4$,  performed  
with  the 
all-electron CC3 method are given in Table~\ref{tab:a2a4cc3}. Extrapolation of these data by 
using the same scheme as for the static polarizability gives
\begin{align*}
 \alpha_2=2.846(5),\;\;\;\alpha_4=4.912(6).\\
\end{align*}
Computation of the post-CC3 and Q corrections to the dispersion coefficients is very complex  and 
is not 
implemented in quantum chemistry software available to us. Therefore, we directly correct for all effects beyond 
the CC3 model by using the FCI calculations in the d-aug-cc-pVDZ basis set using a program written specifically for 
this 
purpose \cite{przybytek14}. Similarly as before, a very conservative accuracy estimate of 50\% is 
used for this quantity, giving $-0.0089(44)$ and $-0.026(13)$ 
contributions to $\alpha_2$ and $\alpha_4$, respectively.

\subsection{Relativistic and QED corrections to the polarizability}

In order to calculate the static polarizability of the neon atom with the accuracy better than a few parts per 
thousand, one has to consider a 
number of corrections accounting for various effects beyond the non-relativistic clamped-nuclei Schr\"{o}dinger 
equation. The list consists of
\begin{enumerate}
 \item relativistic correction of the order $1/c^2$,
 \item leading-order ($1/c^3$) quantum electrodynamics (QED) correction,
 \item finite nuclear mass (FNM) correction,
 \item finite nuclear size (FNS) correction,
 \item higher-order ($1/c^4$ and higher) relativistic and QED effects,
\end{enumerate}
where $c$ denotes the speed of light in vacuum (employed instead of 
the fine-structure constant to avoid a notational collision).

Calculation of the relativistic corrections is based on the Breit-Pauli Hamiltonian~\cite{bethe75}
\begin{align}
 \hat{H}_{\rm BP} = \hat{P}_4 + \hat{D}_1 + \hat{D}_2 + \hat{B},
\end{align}
where the individual operators are defined as
\begin{align}
 \label{p4}
 \hat{P}_4 = -\frac{1}{8c^2}\,\sum_i \nabla_i^4,
\end{align}
\begin{align}
 \label{d1}
 \hat{D}_1 = \frac{\pi}{2c^2} Z\, \sum_i \delta(\textbf{r}_{ia}),
\end{align}
\begin{align}
 \label{d2}
 \hat{D}_2 = \frac{\pi}{c^2} \sum_{i>j}\delta(\textbf{r}_{ij}),
\end{align}
\begin{align}
 \label{bb}
 \hat{B} = \frac{1}{2c^2}\sum_{i>j} \left[\frac{\nabla_i\cdot\nabla_j}{r_{ij}}
+\frac{\textbf{r}_{ij}\cdot(\textbf{r}_{ij}\cdot\nabla_j)\nabla_i}{r_{ij}^3} \right],
\end{align}
where $Z$ is the nuclear charge.
 Relativistic corrections to $\alpha_n$, $n=0,2,4$, due to the operators $X$ of Eqs. (\ref{p4})--(\ref{bb}), will be denoted by 
$\alpha_n^{(2)}(X)$.  
The explicit expression for    $\alpha_n^{(2)}(X)$    can be obtained by  adding $\lambda X$ to the nonrelativistic Hamiltonian, where
 $\lambda$ 
is  a formal parameter, evaluating the polarizability $\alpha_n(X)$,  and extracting terms linear in 
$\lambda$. 

The explicit formulas for  $\alpha_n^{(2)}(X)$ are known~\cite{piszczu15},  but are difficult to implement in practice 
for many-electron systems. 
Therefore,  we adopted a finite-field [for $\alpha_0^{(2)}(X)$] or finite-difference approach [for $\alpha_n^{(2)}(X)$, $n\ge 2]$. 
 In the case of static polarizability, the expectation values of the operators  
from Eqs.~(\ref{p4})--(\ref{bb}) were evaluated 
analytically by the linear response  approach \cite{coriani04} and then numerically differentiated twice with respect to the strength of the applied external electric field. The 
electric-field strengths in the interval~$0.0-0.005$ were tested  and  in all cases at least two 
significant digits were stable 
in the final results. These calculations were performed with the help of the \textsc{Dalton} program package 
using the orbital-relaxed 
CCSD theory with perturbative treatment of triple excitations [CCSD(T)] as described in 
Ref.~\cite{coriani04}. This method of evaluating CC properties is based on the 
so-called CC Lagrangian which, in contrast to the standard CC energy expression, is variational 
with respect to all wavefunction parameters~\cite{adamowicz84,fitz86,salter89,koch90}. This enables 
to obtain first-order properties using a generalization of the Hellmann-Feynman theorem.

For the  dispersion coefficients $\alpha_2$ and~$\alpha_4$, we neglected the two-electron 
relativistic corrections, 
i.e., $\hat{D}_2$ and~$\hat{B}$, since the accuracy requirements are less stringent in this case. Since differentiation 
with respect to 
the external field is not applicable for the dispersion coefficients, a different approach was adopted. The  
operators~$\hat{P}_4$ and~$\hat{D}_1$ multiplied by a formal parameter~$\lambda$ were  added to the Hamiltonian and then 
the dispersion coefficients were calculated analytically using the CC3 level of theory. The values 
of the relativistic 
corrections $\alpha_0^{(2)}(\hat{P}_4)$ and $\alpha_0^{(2)}(\hat{D}_1)$ were extracted  by computing the first 
derivative with respect to $\lambda$ employing   two-point finite-difference formula. The results were sufficiently 
stable for $\lambda$ within the range $5\cdot10^{-5}-1\cdot10^{-3}$. 
 
The one-electron relativistic corrections to the polarizabilities and dispersion coefficients were 
obtained by using a 
modified Slater-type basis sets -- a common set of~$20s15p$ functions (obtained by minimization of the atomic Hartree-Fock energy) 
was used in all basis sets instead of the original sets used in the nonrelativistic calculations. The remaining polarization 
functions were unchanged.

\begin{table*}[t]
\caption{\label{tab:a0rel}
One-electron relativistic corrections to the static polarizability of the neon atom calculated at the CCSD(T) level of 
theory with the finite-field approach.}
\begin{ruledtabular}
\begin{tabular}{ccccccccccccccccc}
basis set & \multicolumn{3}{c}{s-aug} & 
\multicolumn{3}{c}{d-aug} & \multicolumn{3}{c}{t-aug} \\
\hline\\[-1em]
 & $\alpha_0^{(2)}(P_4)$ & $\alpha_0^{(2)}(D_1)$ & sum
 & $\alpha_0^{(2)}(P_4)$ & $\alpha_0^{(2)}(D_1)$ & sum
 & $\alpha_0^{(2)}(P_4)$ & $\alpha_0^{(2)}(D_1)$ & sum \\
\hline\\[-1em]
wtcc2/sp & 0.01826 & $-$0.01465 & 0.00360 & 0.02183 & $-$0.01749 & 0.00434
         & 0.02195 & $-$0.01758 & 0.00436 \\
wtcc3/sp & 0.01748 & $-$0.01395 & 0.00353 & 0.02114 & $-$0.01701 & 0.00413
         & 0.02181 & $-$0.01752 & 0.00429 \\
wtcc4/sp & 0.01893 & $-$0.01521 & 0.00372 & 0.02153 & $-$0.01732 & 0.00422
         & 0.02160 & $-$0.01736 & 0.00423 \\
wtcc5/sp & 0.01965 & $-$0.01580 & 0.00385 & 0.02147 & $-$0.01728 & 0.00420
         & 0.02152 & $-$0.01731 & 0.00421 \\
wtcc6/sp & 0.02004 & $-$0.01613 & 0.00391 & 0.02145 & $-$0.01726 & 0.00419
         & 0.02149 & $-$0.01729 & 0.00420 \\
\end{tabular}
\end{ruledtabular}
\end{table*}

\begin{table*}[t]
\caption{\label{tab:a24rel}
One-electron relativistic corrections to the dispersion coefficients $\alpha_2$ and $\alpha_4$ calculated at the CC3 
level of theory with the finite-difference approach.}
\begin{ruledtabular}
\begin{tabular}{ccccccccccccccccc}
basis set & \multicolumn{3}{c}{s-aug} & 
\multicolumn{3}{c}{d-aug} & \multicolumn{3}{c}{t-aug} \\
\hline\\[-1em]
 & $\alpha_n^{(2)}(P_4)$ & $\alpha_n^{(2)}(D_1)$ & sum
 & $\alpha_n^{(2)}(P_4)$ & $\alpha_n^{(2)}(D_1)$ & sum 
 & $\alpha_n^{(2)}(P_4)$ & $\alpha_n^{(2)}(D_1)$ & sum \\
\hline
\multicolumn{10}{c}{$\alpha_2$} \\
\hline\\[-1.1em]
wtcc2/sp & 0.04789 & $-$0.03461 & 0.01328 & 0.06868 & $-$0.05059 & 0.01809 & 0.06875 & $-$0.05068 & 0.01807 \\
wtcc3/sp & 0.03971 & $-$0.03623 & 0.00348 & 0.06219 & $-$0.04596 & 0.01623 & 0.06699 & $-$0.04942 & 0.01757 \\
wtcc4/sp & 0.05169 & $-$0.03789 & 0.01380 & 0.06275 & $-$0.04810 & 0.01466 & 0.06508 & $-$0.04906 & 0.01602 \\
wtcc5/sp & 0.05453 & $-$0.04147 & 0.01306 & 0.06268 & $-$0.04803 & 0.01466 & 0.06354 & $-$0.04849 & 0.01506 \\
\hline
\multicolumn{10}{c}{$\alpha_4$} \\
\hline\\[-1.1em]
wtcc2/sp & 0.11947 & $-$0.08409 & 0.03538 & 0.19957 & $-$0.14602 & 0.05355 & 0.19769 & $-$0.14520 & 0.05249 \\
wtcc3/sp & 0.13860 & $-$0.10181 & 0.03680 & 0.17670 & $-$0.12892 & 0.04778 & 0.19043 & $-$0.13924 & 0.05119 \\
wtcc4/sp & 0.13331 & $-$0.09515 & 0.03816 & 0.17784 & $-$0.13225 & 0.04559 & 0.18435 & $-$0.13729 & 0.04706 \\
wtcc5/sp & 0.15610 & $-$0.11765 & 0.03844 & 0.17568 & $-$0.13248 & 0.04321 & 0.18216 & $-$0.13528 & 0.04688 \\
\end{tabular}
\end{ruledtabular}
\end{table*}

The computed one-electron relativistic corrections to the static polarizability are shown in Table \ref{tab:a0rel}.
It is seen that
the convergence to   the basis set limit  is rather slow but small basis sets already give accurate results due to 
the presence of the 
common~$20s15p$ functions. Based on the values provided in Table \ref{tab:a0rel}, it can be assumed that 
\begin{align*}
 \alpha_0^{(2)}(P_4) = 0.02149(10),\;\;\;
 \alpha_0^{(2)}(D_1) = -0.01729(10),
\end{align*}
where the uncertainties account for the basis set incompleteness error and for the error due to the finite-field 
approach. Note that there is a significant cancellation between the contributions from the 
$\hat{P}_4$ and $\hat{D}_1$ operators. Because of that the error in the sum of P4 and D1 corrections 
is much smaller than in the individual components. The final one-electron relativistic correction to 
the atomic polarizability of the neon atom is thus equal to~$0.00420(5)$, where error reduction by 
a factor of two with respect to the P4 and D1 corrections was obtained. This value compares quite 
well with the values~$0.00432$, $0.00428$, and $0.00443$ (depending on the basis set) obtained by 
Klopper \emph{et al.}~\cite{klopper04} employing the so called direct perturbation theory~(DPT)  
\cite{Rutkowski86,Kutzelnigg89} at the CCSD(T) level. 

The corresponding results for the dispersion coefficients are given in Table~\ref{tab:a24rel}. Note that the 
convergence of the results is somewhat erratic in basis sets with low augmentation levels. Fortunately, for 
triply-augmented basis sets these problems disappear allowing for the standard $X^{-3}$ extrapolation. One may also 
note 
that the relativistic corrections to the dispersion coefficients are relatively much larger than in 
the case of 
static polarizability. As the final values we take
\begin{align*}
 &\alpha_2^{(2)}(P_4) = 0.0619(16),\;\;\; &&\alpha_2^{(2)}(D_1) = -0.0479(1),\\
 &\alpha_4^{(2)}(P_4) = 0.1799(23),\;\;\; &&\alpha_4^{(2)}(D_1) = -0.1332(21).
\end{align*}
The above results were obtained by $X^{-3}$ extrapolation from the $X=4,5$ pair of basis sets and 
the error was estimated as a difference between the extrapolated result and the value calculated 
with the $X=5$ basis set. The total relativistic correction to the dispersion coefficients 
$\alpha_2$ and $\alpha_4$ equals $\alpha_2^{(2)}=0.0141(10)$ and $\alpha_4^{(2)}=0.0467(1)$, 
respectively. The estimated errors are smaller than of the individual D1 and P4 corrections given above because in 
the present case we directly extrapolated the sum of the P4 and D1 corrections exploiting a systematic error 
cancellation.

\begin{table*}[t]
\caption{\label{tab:a0rel2}
Two-electron relativistic corrections to the static polarizability of the neon atom calculated at the CCSD(T) level of 
theory with the finite-field approach.}
\begin{ruledtabular}
\begin{tabular}{ccccccccccccccccc}
basis set & \multicolumn{3}{c}{s-aug} & 
\multicolumn{3}{c}{d-aug} & \multicolumn{3}{c}{t-aug} \\
\hline\\[-1em]
 & $\alpha_0^{(2)}(D_2)$ & $\alpha_0^{(2)}(B)$ & sum
 & $\alpha_0^{(2)}(D_2)$ & $\alpha_0^{(2)}(B)$ & sum
 & $\alpha_0^{(2)}(D_2)$ & $\alpha_0^{(2)}(B)$ & sum \\
\hline\\[-1em]
cc-pVDZ & 0.000001 & 0.000794 & 0.000794 & 0.000059 & 0.001262 & 0.001321 & 0.000060 & 0.001277 & 0.001337 \\
cc-pVTZ & 0.000037 & 0.001031 & 0.001068 & 0.000063 & 0.001286 & 0.001349 & 0.000063 & 0.001290 & 0.001353 \\
cc-pVQZ & 0.000046 & 0.001175 & 0.001221 & 0.000055 & 0.001281 & 0.001336 & 0.000055 & 0.001281 & 0.001336 \\
cc-pV5Z & 0.000045 & 0.001239 & 0.001284 & 0.000048 & 0.001279 & 0.001327 & 0.000048 & 0.001279 & 0.001327 \\
\end{tabular}
\end{ruledtabular}
\end{table*}

The two-electron relativistic corrections to the static polarizability are shown in Table 
\ref{tab:a0rel2}. They were obtained with help of aug-cc-pV$X$Z Gaussian basis sets,~$X=2-5$ 
\cite{dunning89,kendall92} which were additionally uncontracted in these calculations. In the 
case of the Darwin correction, we apply the~$X^{-1}$ extrapolation formula which gives
\begin{align*}
 \alpha_0^{(2)}(D_2) = 0.00002(1).
\end{align*}
This formula is based on analytic results for the helium atom~\cite{kutz08} and has been used 
successfully in numerous studies for larger systems 
\cite{middendorf12,bischoff10,otto97,przybytek10,przybytek17,cencek12}.
Clearly, this correction is very small and extrapolates to nearly zero. In the case of the Breit correction, we found 
that there is no need to extrapolate the results. The values of~$\alpha_0^{(2)}(B)$ are remarkably stable with 
respect to the size of the basis set, as shown in Table \ref{tab:a0rel2}, and this gives us the following estimate
\begin{align*}
 \alpha_0^{(2)}(B) = 0.00128(1).
\end{align*}
Therefore, the total two-electron relativistic contribution to the polarizability of neon atom amounts to~$0.00130(2)$. 
It seems that there are no literature results for neon that we could confront this value with. 

The data provided above reveal a substantial mutual cancellation of individual contribution to the total relativistic
 corrections $\alpha_n^{(2)}$. 
   For example, the total, i.e., including both one- and 
two-electron contributions, relativistic correction to $\alpha_0$ is about $0.0055$ while the 
largest of these contributions (the mass-velocity term) is $0.02149$. In the case of the dispersion 
coefficients, a similar phenomenon occurs but is somewhat less pronounced. This can be contrasted 
with the analogous data for the helium atom (see Table VIII in Ref.~\cite{puchalski16}) where the 
cancellation is very strong, especially for higher values of $n$.

Next we consider the leading-order quantum electrodynamics (QED) corrections; they shall be denoted as 
$\alpha_n^{(3)}$, 
$n=0,2,4$, further in the text. The correction applied in this work 
reads~\cite{caswell86,pachucki93,pachucki98}
\begin{align}
 \label{hqed}
 \alpha_n^{(3)} = \frac{8}{3\pi c}\left(\frac{19}{30}+2\ln c - \ln k_0\right) \alpha_n^{(2)}(D_1),
\end{align}
where $\ln k_0$ is the so-called Bethe logarithm~\cite{bethe75,schwartz61}. In comparison with the 
full QED treatment that has 
recently been reported for the helium atom~\cite{puchalski20}, formula (\ref{hqed}) contains two 
  simplifications. First, the 
two-electron terms are omitted in~$\hat{H}_{\rm QED}$; this is justified because they are expected to be at least 
several times smaller than the dominant correction~(\ref{hqed}). Second, the electric field dependence of the Bethe 
logarithm is  neglected. 
Calculation of the second electric field derivative of~$\ln k_0$ is very challenging and 
has been accomplished thus far only for the helium atom~\cite{lach04,puchalski20}. 
Moreover,  this derivative (in atomic units) was found to be two 
orders of magnitude smaller than the zero-field Bethe logarithm and thus 
constitutes a negligible 
correction. The reason for this unexpected behavior can be understood by noting that the value of the 
Bethe logarithm is 
sensitive primarily to the region of the wavefunction close to the nucleus. This regime is dominated by the electric 
field generated by the nucleus which is much stronger than the (perturbative) external fields.

The Bethe logarithm for the neon atom without external electric field was calculated at the Hartree-Fock level of 
theory within a basis set of Slater-type orbitals that includes $1p$ functions. Details of these calculations will be 
described in a separate publication. The final result is $\ln k_0=7.595$ and we estimate that this value is accurate to 
within $1-2\%$. This assumption is based on comparisons of analogous results for lighter atoms for 
which more accurate reference data are available. By using the value of the one-electron Darwin 
corrections given earlier, we obtain the following values 
of the QED correction to the static polarizability
\begin{align*}
 \alpha_0^{(3)}=-0.00031(6),
\end{align*}
and to the dispersion coefficients
\begin{align*}
 \alpha_2^{(3)}=-0.00085(17),\;\;\;
 \alpha_4^{(3)}=-0.0024(5),
\end{align*}
where we have assumed an uncertainty of 20\% that accounts for all approximations discussed above. It turns out that 
in the case of the static polarizability, this correction is not negligible within the present 
accuracy standards.

To estimate the contribution of the higher-order  relativistic and QED effects, the so-called one-loop 
correction~\cite{eides01} was calculated. This is 
straightforward because  for atomic systems  this correction  to the polarizability can be expressed  as
\begin{align}
 \alpha_n^{(4)} = \frac{2 Z}{c^2}\, \bigg[\frac{427}{96} - 2 \ln 2\bigg ]\,\alpha_n^{(2)}(D_1),
\end{align}
and thus the computation of this correction amounts to scaling the values of one-electron Darwin corrections given 
above. If one conservatively assumes that the error of neglecting all other higher-order QED diagrams is less than 50\%,
estimations $\alpha_0^{(4)} =-0.00006(3)$ is obtained for the static polarizability, and $\alpha_2^{(4)} =-0.00016(8)$, 
$\alpha_4^{(4)} =-0.00043(21)$ for the dispersion coefficients. This clearly indicates that higher-order relativistic 
and  QED effects are 
negligible from the point of view of the present work and that the perturbative series of QED corrections is rapidly 
convergent.

\subsection{Finite nuclear mass and size corrections to the polarizability}

We also considered the finite nuclear size (FNS) and finite nuclear mass (FNM) corrections, denoted 
$\alpha_n^{\rm FNS}$ and $\alpha_n^{\rm FNM}$ further in the text. For neon atom, the former correction can be obtained 
from the formula
\begin{align}
 \label{a0fns}
 \alpha_n^{\rm FNS} = \frac{4}{3} \frac{\langle r_c^2\rangle}{\lambdabar^2} 
 \,\alpha_n^{(2)}(D_1),
\end{align}
see e.g., Ref.~\cite{puchalski10},
where $\langle r_c^2\rangle$ is the averaged square of the nuclear charge radius and $\lambdabar\approx 386.2\,$ fm is 
the reduced Compton wavelength of the electron. For the $^{20}\mbox{Ne}$ nuclide, the value 
$\langle r_c^2\rangle\approx8.952\,$fm$^2$ 
was taken from the literature \cite{devries87}. This gives $\alpha_0^{\rm FNS}\approx 
1.4\cdot10^{-6}$
revealing that the FNS correction changes the static polarizability of the neon atom by only about 1 ppm and thus it is 
entirely negligible compared to other sources of error. The same conclusion holds for the dispersion 
coefficients.

Moving to the FNM effects, let us consider first the static polarizability. The leading-order FNM correction is 
composed of two terms~\cite{pachucki00}
\begin{align}
\label{a0fnm}
 \alpha_0^{\rm FNM} & = 3\,\frac{m_{\rm e}}{m_N}\,\alpha_0 + 
 \frac{m_{\rm e}}{m_N}\,\partial_{\mathcal{E}}^2\,\langle\sum_{i\neq j}\nabla_i\nabla_j\rangle,
\end{align}
where $m_{\rm e}$ and $m_N$ are the electron and the nuclear masses, respectively, and 
$\partial_{\mathcal{E}}^2\,\langle\sum_{i\neq j}\nabla_i\nabla_j\rangle$ denotes the second electric 
field derivative of the 
expectation value $\langle\sum_{i\neq j}\nabla_i\nabla_j\rangle$ evaluated at zero field 
(mass-polarization term). The first term of 
the above expression (resulting from the reduced-mass scaling) amounts to $0.00022(1)$ 
for $^{20}\mbox{Ne}$. The mass-polarization term is usually about an order of magnitude smaller than the 
mass-scaling term~\cite{pachucki00} and thus can be neglected here. The reduced-mass scaling applied the dispersion 
coefficients  
leads to
\begin{align}
\label{anfnm}
 \alpha_n^{\rm FNM} & \approx (n+3)\,\frac{m_{\rm e}}{m_N}\,\alpha_n.
\end{align}
  Neglecting  the mass-polarization  contribution, we find that  $\alpha_2^{\rm FNM}=0.00039(1)$ and $\alpha_4^{\rm 
FNM}=0.00094(1)$.

\subsection{Final error budget of the polarizability calculations}

\begin{table}[t]
\caption{\label{tab:budget}
Final error budget of the calculations of the static polarizability and dispersion coefficients for 
the neon 
atom.}
\begin{ruledtabular}
\begin{tabular}{llll}
 & \multicolumn{1}{c}{$\alpha_0$} & \multicolumn{1}{c}{$\alpha_2$} & \multicolumn{1}{c}{$\alpha_4$} 
\\
\hline
\multicolumn{4}{c}{clamped-nuclei non-relativistic contributions} \\
\hline
unrelaxed CC3                      &\phantom{$-$}2.66172(18)    &\phantom{$-$}2.846(5) &\phantom{$-$}4.912(6) \\
triples corr.            &\phantom{$-$}0.00034(5)\phantom{1}     & \multicolumn{1}{c}{---} & \multicolumn{1}{c}{---} \\
quadruples corr.         &$-$0.00614(17) & \multicolumn{1}{c}{---} & \multicolumn{1}{c}{---} \\
FCI correction                     &$-$0.00047(24) & $-$0.009(4) & $-$0.026(13) \\
\hline
\multicolumn{4}{c}{other contributions} \\
\hline
one-el. relativistic &\phantom{$-$}0.00420(5) &\phantom{$-$}0.014(1) &\phantom{$-$}0.047(1) \\
two-el. relativistic &\phantom{$-$}0.00130(2) & \multicolumn{1}{c}{---} & \multicolumn{1}{c}{---} \\
leading-order QED        & $-$0.00031(6) &$-$0.001(1) & $-$0.002(1) \\
high-order QED            & $-$0.00006(3) & \phantom{$-$}0.000(1) & \phantom{$-$}0.000(1) \\
finite nuclear mass                &\phantom{$-$}0.00022(1) &\phantom{$-$}0.000(1) &\phantom{$-$}0.001(1) \\
finite nuclear size                &\phantom{$-$}0.00000(1) & \multicolumn{1}{c}{---} & \multicolumn{1}{c}{---} 
\\
\hline
total                 &\phantom{$-$}2.66080(36)  &\phantom{$-$}2.850(7) &\phantom{$-$}4.932(14) \\
\vspace{-0.5cm}
\end{tabular}
\end{ruledtabular}
\end{table}

The final theoretical results for the static polarizability and dispersion 
coefficients of the neon atom are summarized in Table~\ref{tab:budget}. We show the 
magnitude of various contributions to $\alpha_n$ along with their error estimation, according to 
the discussion from the previous sections. The final error is obtained by summing the errors in the individual 
components quadratically. Clearly, the dominant source of error is the 
insufficiently accurate description of higher-order excitations, particularly of the FCI correction. In 
fact, with our resources it would still probably be possible (albeit at a huge cost) to perform CC3 
calculations with the cardinal number $X$ of the basis set size increased by one. Coupled with the CBS 
extrapolation, which was shown to perform quite well at the CC3 level of theory, this may lead to 
the  error 
reduction by a factor of about two in the CC3 component of the polarizability. However, this would 
not decrease the overall uncertainty of our result since already at the present level it is
dominated by the error of the FCI correction.

Rather surprisingly, the relativistic corrections to $\alpha_n$ can be calculated with a sufficient 
accuracy. While they are extremely complicated from the point of view of analytical evaluation, it 
appears that the mixed analytic and finite-field approach adopted by us is adequate. It is also 
worth mentioning that the approximations adopted by us in the computations of the QED corrections do 
not contribute significantly to the overall error. The errors resulting from omission of the 
two-electron QED contributions and from missing electric-field derivative of the Bethe logarithm 
would become important only when other sources of error were reduced by a factor of at least five.

The data provided in Table~\ref{tab:budget} also reveal  a substantial cancellation between 
various contributions to the static polarizability. Indeed, the one- and two-electron relativistic 
corrections combined amount to about $0.0053$ and are of opposite sign to the total nonrelativistic 
post-CC3 correction which is equal to $-0.0063$. This means that the difference between the pure 
CC3 result and the experimental vale is much smaller than one could expect \emph{a priori}. The 
same effect is present also in calculations of the dispersion coefficients, albeit to a smaller 
extent. This may also explain an unexpectedly good agreement of some literature values with the 
experimental results despite the fact that all post-CC3/post-CCSD(T) and relativistic corrections were neglected.

\subsection{Magnetic susceptibility}

\begin{table}[t]
\caption{\label{tab:r2}
Expectation values $\langle \Sigma_i r^2_i\rangle$ calculated for the neon atom. The valence and core 
correlation contributions were obtained at the CC3/(aug-)cc-pVXZ and CC3/(aug-)cc-pCVXZ levels of 
theory, respectively. }
\begin{ruledtabular}
\begin{tabular}{cccccc}
       & DZ & TZ & QZ & 5Z & 6Z \\
\hline
\multicolumn{6}{c}{valence contribution} \\
\hline
no-aug & 9.1422 & 9.3894 & 9.4929 & 9.5544 & 9.5661 \\
aug    & 9.7022 & 9.6336 & 9.5994 & 9.5802 & 9.5740 \\
\hline
\multicolumn{6}{c}{core correlation contribution} \\
\hline
no-aug & $-$0.0017 & $-$0.0062 & $-$0.0079 & $-$0.0088 & --- \\
aug    & $-$0.0026 & $-$0.0071 & $-$0.0085 & $-$0.0089 & --- \\
\end{tabular}
\end{ruledtabular}
\end{table}

As mentioned in Sec. \ref{sec:intro}, the determination of temperature or pressure via measurements 
 of the refractive index $n$ 
 requires also the knowledge of the magnetic 
susceptibility, $\chi$. For the neon atom this quantity is about $3\cdot 10^{-5}$ times smaller in magnitude than the 
static polarizability \cite{robert2019crc}. Therefore,   if  the required  relative 
accuracy in calculation of the refractive index is of the order of $10^{-6}$, it is  
sufficient to know $\chi$   with an uncertainty  of only 1 percent, as evident from the 
Clausius-Mossotti relations for $\epsilon_r$ and $\mu_r$. This allows us to adopt   significant  simplifications in 
evaluation of the magnetic susceptibility, 
namely (i) the 
frequency dependence of $\chi$ can be entirely neglected  and (ii) the static value of 
$\chi_0$ can be calculated from the ``non-relativistic'' formula \cite{bethe75},
\begin{align}
\label{chieq}
 \chi_0=-\frac{e^2}{6m_{\rm e}\,c^2}\,\big< \sum_i r_i^2\big>,
\end{align}
where $\langle   r_i^2\rangle$ is the mean square distance of an  electron  from the nucleus.

The quantity $\langle \sum_i r_i^2\rangle$ was computed directly as an expectation value with the 
CC3 wavefunction using the formalism proposed by Tucholska~\emph{et al.}~\cite{tucholska14} In this 
method one introduces an auxiliary excitation operator $S$~\cite{jeziorski93}, defined by a 
closed-form linear equation involving the standard cluster operator $T$ and its Hermitian conjugate. 
With help of the $S$ operator one can rewrite the CC expectation value of an arbitrary operator as a 
finite commutator expansion. Since the exact $S$ operator corresponding to the CC3 wavefunction 
involves excitation levels higher than triple, it is necessary to truncate it for practical 
reasons. In this work we employ the truncation at the third-order of perturbation theory as 
suggested in Ref.~\cite{tucholska14} which provides an optimal balance between the accuracy and 
computational costs.

The valence results, i.e., with the frozen $1s^2$ core, were obtained at the CC3/(aug-)cc-pVXZ 
level of theory. For the evaluation of the core correlation contribution we used the extended 
(aug-)cc-pCVXZ basis sets that include additional tight functions  (with large exponents) 
for a better description of the core region. The results are shown in Table \ref{tab:r2}.  Both the 
valence and core contributions were 
extrapolated to the complete basis set limit from the largest two augmented basis sets
using the $X^{-3}$ formula. In each case, we determine the error as a difference between the 
extrapolated and the largest basis set results. This gives us $9.565(9)$ and $-0.009(1)$ for the 
valence and core correlation contributions, respectively. Concerning the augmentation level, we 
found the single augmentation to be entirely sufficient for the calculation of $\langle \sum_i 
r_i^2\rangle$. We checked that already at the 
quadruple-zeta level, the inclusion of the second set of diffuse functions changes the results shown 
in Table \ref{tab:r2} only at the last significant digit. Finally, we add the FCI correction that 
we managed to compute only with  aug-cc-pVDZ and aug-cc-pVTZ basis sets, obtaining $0.00148$ and 
$0.00170$, respectively. By extrapolating these two values using the $X^{-3}$ formula, we obtain our 
final estimation of the FCI correction, equal to $0.0018(1)$. By combining the valence, 
core correlation, and FCI contributions and adding the errors quadratically we find
\begin{align}
 \langle \sum_i r_i^2\rangle = 9.558(10),
\end{align}
which is our final value of the mean square distance of the electrons from the nucleus. The ``nonrelativistic'' 
diamagnetic susceptibility of the neon atom is thus
\begin{align}
\label{chival}
\begin{split}
 \chi_0 &= - 8.483(8)\cdot10^{-5}\; \mbox{a.u.}\\
\end{split}
\end{align}
or   $\chi_0  = -7.570(8)\cdot10^{-6}\; \mbox{cm}^3\mbox{/mol}$ 
in units used conventionally in experimental work.

Below we consider corrections to this result that account for effects not included in the 
``nonrelativistic'' approximation of Eq. (\ref{chieq}).  The  finite-nuclear-mass correction,  
$\chi_0^{\rm FNM}$, to the magnetic susceptibility  was considered 
  by Bruch and Weinhold~\cite{bruch02} for the helium atom, see also Ref. \cite{bruch03} for 
an erratum to this work. 
These authors derived the complete formula, including the mass-polarization term,  for the leading  
contribution to $\chi_0^{\rm FNM}$  of the order  of  $ m_{\rm e}/m_{\rm N}$.   They found that 
the mass-polarization  correction is by about 5 orders of magnitude  smaller than $\chi_0$. 
For the neon atom, this ratio will be even smaller because of much larger mass  of the 
$^{20}$Ne nucleus. Therefore, in our work  we neglected the mass-polarization contribution to 
$\chi_0$.  The derivation of Bruch and Weinhold was recently generalized to   many-electron
atoms by  Pachucki and Yerokhin \cite{pachucki19}.  When the mass-polarization term is neglected, 
their formula reads 
\begin{align}
\label{chifnm}
 \chi_0^{\rm FNM} = -\frac{e^2}{2m_N\,c^2}\,\big< \sum_i r_i^2\big> -\frac{e^2}{6m_N\,c^2}\,
  \big<\sum_{i\neq j}\mathbf{r}_i\cdot\mathbf{r}_j \big>.
\end{align}
The first term is simply a scaling of the value  of Eq.~(\ref{chival}) by a factor of $3m_{\rm 
e}/m_N$. This 
gives a 
correction equal to  $-6.98\cdot10^{-9}\,$a.u. which is not entirely  negligible in the present 
context. 

Bruch and Weinhold \cite{bruch02,bruch03} found that the second term in 
Eq. (\ref{chifnm}), i.e., involving the operator
$\sum_{i\neq j}\mathbf{r}_i\cdot\mathbf{r}_j$, 
 is very small, even smaller than the mass-polarization term.
  However, unlike for helium, in the case of  
neon  the expectation value $\langle\sum_{i\neq j}\mathbf{r}_i\cdot\mathbf{r}_j\rangle$ does 
not vanish at the uncorrelated, i.e., at  the Hartree-Fock level, due to the presence of the $p$ 
orbitals 
in the ground-state reference determinant. Therefore, the estimation  based on the helium results 
may 
not be reliable and we decided to calculate this term in an approximate way to remove this  
potential source of uncertainty. For this purpose we used the CC3 wavefunction and calculated the 
two-electron expectation value in the leading order of the perturbation theory~\cite{jeziorski93}. 
Using the aug-cc-pCV5Z basis set and with all electrons correlated, we found that 
\begin{align}
 \langle\sum_{i\neq j}\mathbf{r}_i\cdot\mathbf{r}_j\rangle = -3.614.
\end{align}
This expectation value turns out to be quite large, only about two  and a half times smaller than 
the one-electron 
counterpart, $\langle \sum_i r_i^2\rangle$. Still, the second term in Eq.~(\ref{chifnm}) amounts to 
only 
$8.8\cdot10^{-10}\,$a.u. and can be neglected even in computing the total 
uncertainty 
of  $\chi$. The inclusion of 
the finite nuclear mass corrections in calculation of the  magnetic susceptibility of the neon 
atom  will  be  necessary  only if one aims at achieving the  accuracy level of $10^{-4}$ or better.

Let us finally consider the relativistic correction
 to  $\chi_0$. The leading term in this correction is of the order of  $1/c^4$  and will be denoted 
 by $\chi_0^{(4)}$. 
 Complete computation of $\chi_0^{(4)}$ for   
 a noble gas atom represents a considerable  challenge  and has not  been performed thus far even 
for helium.    
In their helium study,   Bruch and Weinhold~\cite{bruch02} neglected the magnetic-field dependent 
terms  in the Breit-Pauli 
Hamiltonian \cite{bruch03,pachucki03} and  found that the  computed approximation to  
$\chi_0^{(4)}$  is by a factor of 
8$\cdot 10^{-5 }$ smaller  than  nonrelativistic value   of $\chi_0$ and, therefore, negligible for 
this system.  Since the  ratio  $\chi_0^{(4)}/\chi_0$  
  scales quadratically with an effective nuclear charge $Z_{\rm eff}$, the  significance  of    
$\chi_0^{(4)}$ must certainly be much larger  for neon then for helium.  Any choice of  $Z_{\rm 
eff}$ would be to  a large extent arbitrary, so we  decided to  estimate  $\chi_0^{(4)}$
assuming that  the significance of the  relativistic correction to   $\chi_0$ is percentagewise the 
same as in the case of the relativistic correction to $\alpha_0$.  
From Table IX we see that $\alpha_0^{(2)} = 0.00550(6)$  represents  0.2\% \ of  $\alpha_0 $.  
Assuming the same proportion for 
the magnetic susceptibility, we find that   that the magnitude of  $\chi_0^{(4)}$ can be estimated 
as 
$0.017\cdot10^{-5}$  and we add this value 
to the final error budget of $\chi_0$.  This is sufficient for the present  purpose but it is clear 
that the complete calculation of 
the relativistic corrections to the magnetic susceptibility of helium and other noble gases 
represents  interesting topic for a future 
study. 

\section{Discussion and conclusions}

In Table~\ref{tab:literature}, we compare the results  of polarizability calculations  
with other theoretical work and experimental values taken from the literature. While our 
intention was to include the most recent and representative results available currently, we by no 
means claim this list to be exhaustive.
\begin{table}[t]
\caption{\label{tab:literature}
Comparison with other theoretical and experimental literature values of $\alpha_n$. Wherever no 
error estimation is present it means that they were not given by the authors. All values are given 
in the atomic units.}
\begin{ruledtabular}
\begin{tabular}{lD{.}{.}{1.8}D{.}{.}{1.5}D{.}{.}{1.6}}
 & \multicolumn{1}{c}{$\alpha_0$} & \multicolumn{1}{c}{$\alpha_2$} & \multicolumn{1}{c}{$\alpha_4$} 
\\
\hline
\multicolumn{4}{c}{experimental} \\
\hline
Chan et al. \cite{chan92}           & \multicolumn{1}{c}{---} & 2.938 & 5.137 \\
Kumar and Meath \cite{kumar10}      & 2.669 & 2.875 & 4.994 \\
Orcutt and Cole \cite{orcutt67}     & 2.658 & \multicolumn{1}{c}{---} & \multicolumn{1}{c}{---} \\
                                    & 2.663\textsuperscript{a} & \multicolumn{1}{c}{---} & 
\multicolumn{1}{c}{---} \\
Gaiser and Fallmuth \cite{gaiser10} & 2.66110(3) & 
\multicolumn{1}{c}{---} & \multicolumn{1}{c}{---} \\
Gaiser and Fallmuth \cite{Gaiser18} & 2.661057(7) & 
\multicolumn{1}{c}{---} & \multicolumn{1}{c}{---} \\
\hline
\multicolumn{4}{c}{theoretical} \\
\hline
Paw\l{}owski et al. \cite{pawlowski05} & 2.665   & 2.859 &  4.946 \\
Klopper et al. \cite{klopper04}        & 2.66312 & \multicolumn{1}{c}{---}   
& \multicolumn{1}{c}{---} \\
Larsen et al. \cite{larsen99}          & 2.673   & \multicolumn{1}{c}{---}   
& \multicolumn{1}{c}{---} \\
\hline
\multicolumn{4}{c}{this work} \\
\hline
this work                 & 2.66080(36)  & 2.850(7) & 4.932(14) \\
\vspace{-0.5cm}
\end{tabular}
\end{ruledtabular}
\vspace{-0.35cm}
\begin{flushleft}
\small \textsuperscript{a}after applying the correction for the compressibility  
\end{flushleft}
\end{table}

Our final theoretical value of the static polarizability $\alpha_0=2.66080(36)$  has an 
estimated uncertainty of about one part per 10$^4$. The most recent experimental value for this 
quantity, $\alpha_0=2.661057(7)$, obtained by Gaiser and Fellmuth  \cite{Gaiser18}, lies within our 
error 
bars, but is about $50$ times more accurate. Therefore, unlike for helium, the present-day 
theoretical methods for ten-electron atoms are not competitive in terms of accuracy 
 with the  results obtained from the dielectric-constant thermometry experiments. 
Nonetheless, 
it is clear from Table~\ref{tab:literature} that our results are the most reliable theoretical data 
available 
for 
$\alpha_0$, especially taking into account the systematic inclusion of various small 
physical effects and more rigorous error estimations. The best previous theoretical estimate 
for $\alpha_0$ presented  in Ref. \cite{Gaiser18} is obtained by combining the relativistic results 
of Klopper et al. \cite{klopper04} with the FCI correction calculated by Larsen et al. 
\cite{larsen99} resulting in  $\alpha_0=2.6617(20)$. The estimated uncertainty of this value is 
about six times larger than  the uncertainty of our result. One may note that the largest 
contribution to our error budget 
comes from the approximate (small basis set) FCI treatment  of   $n$-electron excitations, $n>4$. 
A significant reduction of this error will be difficult because of the  exponential  scaling of the 
time and memory resources needed to perform  FCI calculations.   

The reliability of the experimental results is less impressive in the case of the dispersion 
coefficients.  The older data of  Chan et al.~\cite{chan92} are significantly less accurate than 
our values. 
The most recent data come from the work  of Kumar and Thakkar \cite{kumar10} who 
employed a semiempirical dipole oscillator strength distribution (DOSD) technique to extract 
$\alpha_n$ from the experimental photoabsorption cross sections with addition of some theoretical 
constraints. These  authors estimated that their results for $\alpha_2$ and $\alpha_4$ are accurate 
to 
about $\pm1\%$ and, therefore, agree to within the combined uncertainty estimates with the values 
calculated here. Similarly, our results also agree well with 
the theoretical data of Paw\l{}owski et al. \cite{pawlowski05} who also estimate a similar accuracy 
level. Note that the dispersion coefficients calculated in the present work [$\alpha_2=2.850(7)$ 
and $\alpha_4=4.932(14)$] are estimated to be accurate to about one-two parts per thousand. 
Therefore, they are by almost an order of magnitude more accurate than the best previous 
estimates, including both theoretical and experimental literature data.

The value of the static magnetic susceptibility $\chi_0$ obtained by us is $-8.484(19)\cdot 
10^{-5}$ 
a.u.,
or   $  -7.571(17)\cdot10^{-6}\; \mbox{cm}^3\mbox{/mol}$, where the reported uncertainties  
include now the entire value of the estimated relativistic correction.   Our result, estimated to 
be 
accurate to about  2 parts per
 thousand, agrees quite well with the older experimental value reported by Havens~\cite{havens33} 
equal to 
 $-7.651\cdot10^{-6}\; \mbox{cm}^3\mbox{/mol}$, but is about  9\% larger in absolute value  than  
the more recent recommended  experimental result  of   $ 
 -6.96\cdot10^{-6}\; \mbox{cm}^3\mbox{/mol}$ \cite{barter60,robert2019crc}
One may note  that also for helium there is about 7\% \ discrepancy between the best theoretical 
result  \cite{bruch02} 
and the recommended  experimental value \cite{robert2019crc}.  The reasons behind these 
differences 
between theory and experiment  
 are not clear. In the case of neon, one can  consider  the following hypothetical  sources of the 
observed discrepancy: \vspace{-1ex} 
\begin{itemize}
 \item [$(a)$]   significant underestimation of relativistic correction to $\chi_0$,   \vspace{-1ex} 
  \item [$(b)$]   neglect of the paramagnetic contribution to  $\chi_0$, \vspace{-1ex} 
 \item  [$(c)$]  neglect of the temperature dependence of $\chi_0$,   \vspace{-1ex} 
 \item [$(c)$]   neglect of the density dependence of $\chi_0$.   \vspace{-1ex} 
\end{itemize}
 The source $(a)$ is unlikely since the relativistic  correction included in our  uncertainty 
estimate represents only  
about 0.2\% \ of $\chi_0$. This value is approximately 25 times larger than the estimate obtained
for helium, suggesting that the ratio $Z_\mathrm{eff}(\mathrm{Ne})/Z_\mathrm{eff}(\mathrm{He})$ is about 5.
Even if this estimation is multiplied by a factor of 4, which corresponds 
to assuming the maximal possible value of the ratio $Z_\mathrm{eff}(\mathrm{Ne})/Z_\mathrm{eff}(\mathrm{He})=10$,
the experimental value 
remains 10$\sigma$ away from our theoretical result.    
  
Regarding point $(b)$, we found that the  leading  nonrelativistic  paramagnetic contribution  to  
$\chi_0$ is of the 
order 
of $(m_{\rm e}/m_{\rm N})^2$  and is therefore entirely negligible. This  contribution was 
considered  in the Appendix of Ref. \cite{bruch02} and incorrectly claimed  to vanish for helium.  
The argument  given  in this Appendix is based on the   
assumption that quantum states of the  helium atom cannot have $^1P^{\rm e}$ symmetry, which is 
incorrect in the continuous spectrum.   
There is  also relativistic paramagnetic contribution  to $\chi_0$ resulting from the relativistic 
$^3P_0$ component of the 
ground-state helium wave function.  This contribution is of the order of $1/c^6$ and is completely 
negligible.  One may note here
that the frequency dependence of $\chi$ may result only from  paramagnetic terms. The arguments 
given above show that 
this frequency dependence  is extremely weak for rare gases and entirely negligible in practice.   

Regarding point $(c)$, one can observe that for an atom moving in the magnetic field, the center of 
mass cannot be 
separated 
and the magnetic susceptibility depends on the momentum  of  the atom.  This effect  was considered 
by Bruch and Weinhold 
in Ref. \cite{bruch02} and found to be four orders of magnitude smaller than  the finite nuclear  
mass correction 
 $\chi_0^{\rm FNM}$ (for  $^3$He at $T \approx 10$ kelvin).  For neon this correction will be  even smaller because of  
its greater mass and 
can be assumed negligible at normal temperatures. 

The point $(d)$, that is the density dependence, was considered by Bruch and Weinhold \cite{bruch00} 
for liquid helium.
 These authors used a simple Hartree-Fock plus dispersion model  for the interaction-induced magnetic susceptibility 
of helium and found that the effect of the interatomic interactions is small and the measurement in the liquid  can be 
used 
to accurately determine  the isolated atom magnetic susceptibility.  This conclusion must obviously apply also  to the gas phase.   
Since the measurement for neon was made in the gas phase, the density dependence of the magnetic susceptibility cannot 
explain the observed  difference between our theoretical value and the experimental one.  The source of this  difference remains 
unclear  to us and represents, as for helium,  a troubling puzzle  to be hopefully resolved in the future by   new theoretical ideas or 
new experiments.

To sum up, we have reported state-of-the-art theoretical determination of the static and dynamic 
polarizability of the neon atom. Numerous corrections beyond the nonrelativistic clamped-nuclei 
picture, such as those due to the quantum electrodynamics effects, have been included for the first time 
for this system and the sources of error have been carefully  discussed. Additionally, we determined the 
 magnetic susceptibility of the neon atom with an accuracy of a few parts per thousand. Since the 
magnetic contribution to the refractive index of the gaseous neon is about five orders of magnitude 
smaller than the very accurately known electric contribution, our result is entirely sufficient to calculate the 
refractive index with an accuracy of one part per million.

\begin{acknowledgments}
We thank Krzysztof Pachucki for discussions and Christof Gaiser, Roberto Gavioso, and Krzysztof 
Szalewicz for comments on the manuscript.
This project (QuantumPascal project 18SIB04) has received funding from the EMPIR programme 
cofinanced by the Participating States and from the European Union’s Horizon 2020 research
and innovation program. The authors also acknowledge support from the National
Science Center, Poland, within the Project No. 2017/27/B/ST4/02739. 
ML was supported by the Foundation for Polish Science (FNP) and by the Polish National Agency 
of Academic Exchange through the Bekker programme No. PPN/BEK/2019/1/00315/U/00001.
\end{acknowledgments}

\bibliography{ne_polar}

\end{document}